\documentclass[aip,cha]{revtex4-1}
\usepackage{amsmath}
\usepackage{graphicx}
\usepackage{color}
\usepackage{textcase}
\usepackage{url}
\usepackage{bm}
\usepackage{ulem}
\renewcommand{\theenumi}{\Roman{enumi}}
\draft 

\begin{document}


\title{Boundary-Induced Pattern Formation from Uniform Temporal Oscillation}



\author{Takahiro Kohsokabe}
\email[]{kohso@complex.c.u-tokyo.ac.jp}
\affiliation{Department of Basic Science, Graduate School of Arts and Sciences, The University of Tokyo - 3-8-1 Komaba, Meguro, Tokyo 153-8902, Japan}

\author{Kunihiko Kaneko}
\email[]{kaneko@complex.c.u-tokyo.ac.jp}
\affiliation{Research Center for Complex Systems Biology, Graduate School of Arts and Sciences, The University of Tokyo - 3-8-1 Komaba, Meguro, Tokyo 153-8902, Japan}
\affiliation{Department of Basic Science, Graduate School of Arts and Sciences, The University of Tokyo - 3-8-1 Komaba, Meguro, Tokyo 153-8902, Japan}


\date{\today}

\begin{abstract}
Pattern dynamics triggered by fixing a boundary is investigated. By considering a reaction-diffusion equation that has a unique spatially-uniform and limit cycle attractor under a periodic or Neumann boundary condition, and then by choosing a fixed boundary condition, we found three novel phases depending on the ratio of diffusion constants of activator to inhibitor: transformation of temporally periodic oscillation into a spatially-periodic fixed pattern, travelling wave emitted from the boundary, and aperiodic spatiotemporal dynamics.  The transformation into a fixed, periodic pattern is analyzed by crossing of local nullclines at each spatial point, shifted by diffusion terms.  A spatial map, then, is introduced, whose temporal sequence can reproduce the spatially periodic pattern, by replacing the time with space. The generality of the boundary-induced pattern formation as well as its relevance to biological morphogenesis is discussed.
\end{abstract}


\maketitle 

\textbf{In spite of extensive studies on the initial-condition dependence, boundary-condition dependence is not fully explored, whereas most pattern dynamics in nature generally progress in a finite system, thus pattern formation following the boundary condition is of crucial importance. We study a system in which temporally-periodic and spatially-uniform attractor is reached in a bulk, and how it is changed by imposing a fixed boundary condition. Besides the trivial case in which the original uniform oscillation is recovered far distant from the boundary, we uncovered non-trivial pattern dynamics that replace the original uniform state globally. In particular, the transformation of oscillation into a fixed spatial pattern is analyzed in depth, which should be essential to biological morphogenesis, such as somitogenesis.}

\section{Introduction}
Pattern formation in nonlinear-nonequilibrium systems has gathered much attention over decades both theoretically and experimentally. With the motivation to understand biological morphogeneis, Turing\cite{turing1952chemical}, in his seminal paper, showed how spatial pattern is spontaneously organized from a spatially homogeneous state, without any external signal, as a result of symmetry breaking.  One of the six possible pattern formation processes he classified is now known as the \textit{Turing pattern}. Stationary periodic pattern with a finite wavelength is self-organized from a temporally stationary and spatially uniform state. 

Since the pioneering study by Turing, spatiotemporal dynamics in chemical-reaction systems have been extensively investigated, including waves and spirals \cite{kuramoto1984chemical}, as well as spatiotemporal chaos\cite{cross1993pattern}. 
On the other hand, understanding morphogenesis according to the Turing's thesis needed more time to be accepted in biology.

The Turing pattern observed so far in biology, however, is still restricted in few cases such as the patterns of shell\cite{meinhardt1982models} or fish skin\cite{kondo1995reaction}. In contrast, relevance of oscillatory dynamics to morphogenesis is now recognized by the discovery of oscillatory gene expressions in the somitogenesis of vertebrate development, and stem cell systems\cite{kobayashi2009cyclic, suzuki2011oscillatory}. In chick development, temporal oscillation in the gene expression of c-hairy1 is uncovered in the presomitic mesoderm. Such temporal oscillation is later fixed into a spatial pattern \cite{palmeirim1998uncoupling,pourquie2001vertebrate}. For fixation into a spatial pattern, relevance of external gradient was discussed as clock-and-wave-front mechanism\cite{cooke1976clock,meinhardt1982models}.

In contrast, we recently found an alternative mechanism for the transformation from temporal oscillation to fixed spatial pattern, induced by diffusive interaction and boundary condition, even without any input of external gradient \cite{kohsokabe2016evolution, kohsokabe2017boundary}. This finding can provide possible explanation for recent experimental results on somitogenesis, where cell-cell interaction is also suggested to be relevant \cite{dias2014somites, cotterell2015local}. As the proposed mechanism of ours requires just a fixed boundary and diffusion of chemicals, we expect that it can be applicable to a variety of chemical and biological systems. 

Here, we first confirm the mechanism by investigating reaction-diffusion systems, focusing on the influence of boundary conditions to pattern formation. For the present study we consider a system in which spatially uniform and temporally oscillatory state exists as an attractor if a boundary condition allows, and then adopt a fixed boundary condition. In this case, temporal oscillation is ceased at the boundary point. Now, for some range of diffusion constants, homogeneous oscillation is recovered as the spatial point is departed from the boundary. In other cases, however, which we will mainly discuss in the present paper, the temporal oscillation is replaced by a spatially periodic pattern, induced by the fixed boundary condition. Note that the boundary sometime can influence on the global pattern, as was discussed in Ref.\cite{pomeau1981wavelength, fujimoto2001sensitive}, while selection of periodic pattern or other spatiotemporal pattern from a homogeneous periodic state is only recently found and will be analyzed here.

After introducing two-component reaction-diffusion equations in \S 2, we study the pattern formation induced by a fixed boundary condition in \S 3. Depending on the diffusion constants, we found fixation into a spatially periodic pattern, traveling wave, and spatiotemporal aperiodic dynamics. The wavelength of  periodic pattern is not obtained by linear stability analysis as in the Turing pattern. Instead, in \S 4, we introduce local nullcline analysis that also includes the effect of diffusion. A four-dimensional spatial map is then introduced in \S 5, which successively determines the fixed pattern from the upper- to down- stream. The periodic attractor of the spatial map determines the organized spatial pattern. Relevance of the present boundary-induced pattern formation to biological problems is discussed in \S 6.

\section{Reaction-Diffusion Model}

We consider a spatially reaction-diffusion system of two components $X(r,t)$ and $Y(r,t)$ on a one dimensional space $r$.
\begin{equation}
\left.
\begin{array}{l}
\dfrac{\partial X}{\partial t}=f(X,Y)+D_X\dfrac{{\partial}^2 X}{\partial r^2}\\[8pt]
\dfrac{\partial Y}{\partial t}=g(X,Y) +D_Y\dfrac{{\partial}^2 Y}{\partial r^2}\\
\end{array}
\right.
\label{eq:reaction-diffusion}
\end{equation}
where $f(X,Y)$ and $g(X,Y)$ are the reaction functions for $X$ and $Y$, $D_X$ ($D_Y$) is the diffusion constant of $X$ ($Y$), respectively. In order to study the fixation into spatial pattern from a homogeneous oscillatory state, we assume that the attractor of the dynamical system without the diffusion term, i.e., the dynamical system $dX/dt=f(X,Y),dY/dt=g(X,Y)$ is a unique limit cycle.
It is generated by a Hopf bifurcation from the fixed point $(X^*,Y^*)$. Hence the Jacobi matrix at the fixed point
\begin{equation}
\begin{pmatrix}
a & b \\ c &d
\end{pmatrix}
=
\begin{pmatrix}
\frac{\partial f}{\partial X} & \frac{\partial f}{\partial Y}\\
\frac{\partial g}{\partial X} & \frac{\partial g}{\partial Y}
\end{pmatrix}_{X=X^*, Y=Y^*}
\label{eq:Jacobian}
\end{equation}
satisfies $a+d>0$ and $-2<\frac{d-a}{\sqrt{-bc}}<2$\cite{turing1952chemical}.
In all the examples to be studied here, the spatially-uniform limit cycle is an attractor as long as the boundary condition allows for its existence. For example, by taking Neumann or periodic boundary conditions, the uniform limit cycle state is the only attractor in the reaction-diffusion equation (1). Throughout the present paper, we discuss such case and how the boundary condition alters an attractor.

To be specific we study the following two systems.
The first one (model A) is a simplified from of the gene regulation network, while the second one is Brusselator (model B). 
The first one is given by 

{\bf Model A}
\begin{equation}
\left.
\begin{array}{l}
f(X,Y)=\dfrac{1}{1+e^{-\beta (Y-1/2)}}-X\\[12pt]
g(X,Y)=\dfrac{1}{1+e^{-\beta (Y-X)}}-Y.\\
\end{array}
\right.
\label{eq:example1}
\end{equation}
where $X$ and $Y$ are the protein concentrations. $X$ inhibits the expression of the protein $Y$ while $Y$ activates the expressions of both $X$ and $Y$\cite{mjolsness1991connectionist, PhysRevE.88.032718}.
When $\beta>8$, this system has one unstable fixed point $(X^*,Y^*)=(1/2,1/2)$, and the limit cycle is an attractor.  Unless otherwise mentioned, we choose $\beta=40$ throughout the paper.

The second one, the so-called Brusselator is given by Ref.\cite{nicolis1977self}: 

{\bf Model B}
\begin{equation}
\begin{array}{l}
f(X,Y)=BY-XY^2\\[8pt]
g(X,Y)=A-(B+1)X+XY^2\\
\end{array}
\label{eq:brusselator}
\end{equation}
As is well known, the systems has a limit cycle attractor for $A^2+1<B<(A+1)^2$.  Here we choose $A=1$ and $B=3$.

\section{Pattern Formation by Boundary Condition}

Now we adopt a fixed boundary condition $X(0,t)=X_0$ and $Y(0,t)=Y_0$ for the above reaction-diffusion models. The fixed boundary is set at the leftmost point $r=0$, and we study how this choice alters an attractor by taking the models (A) and (B). In order to exclude the effect of initial condition to pattern formation dynamics, the uniform initial condition with the boundary values, i.e., $X(r,0)=X_0$ and $Y(r,0)=Y_0$ is adopted. Unless otherwise mentioned the right boundary is given by Neumann condition, so that the uniform state is not destabilized from that boundary.  Now by the fixed boundary at the left side $r=0$, we found that the spatially uniform oscillation is not always reached, even if the system goes sufficiently far to the right side (which we call downflow). 

Depending on the values of diffusion constants $D_X$ and $D_Y$, the behaviors downflow are characterized into the following phases.

\begin{enumerate}
	\item The oscillation downflow is replaced by a temporally-fixed and spatially-periodic pattern.

	\item As the spatial location goes downflow, $X$ and $Y$ oscillate in time. The amplitudes of oscillation increase with $r$. As $r$ goes larger, the original spatially-uniform, limit cycle attractor is reached.

	\item Neither spatially-uniform nor temporally-fixed pattern is reached. Instead, travelling wave is emitted from the boundary, which replaces the original uniform oscillatory state.

	\item Neither temporally periodic nor spatially periodic state is stable. Spatiotemporal aperiodic dynamics replaces the original uniform oscillatory state.
\end{enumerate}

By using the model (A), the typical spatiotemporal patterns in the four phases are shown in Fig.1, which appear depending on the diffusion constants $D_X$ and $D_Y$. The phase diagram against the diffusion constants is given in Fig.2, where the phase is determined basically by $D_Y/D_X$. The parameter region for each phase is (I) $D_Y/D_X <0.15$ (II) $ 0.15<D_Y/D_X <0.35$ (III) $0.35< D_Y/D_X <1.6$ (IV) $1.6< D_Y/D_X $. (Fig.1a-c) belongs to the phase (I), (d) to (II), (e) to (III), and (f) to (IV).

\begin{figure}
\includegraphics{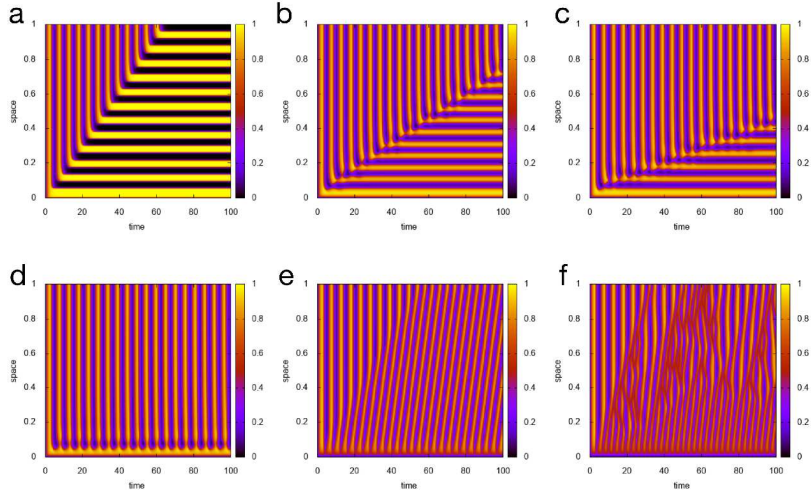}
\caption{Time development of the model \textbf{A} under a fixed boundary condition, $X_0=Y_0=0$. $X(r,t)$ is displayed with a color scale given by the side bar. The abscissa is space $r$, and the ordinate is time $t$. $D_X$ is set to $5\times10^{-4}$ throughout the subfigures (a)-(f). Unless otherwise mentioned, the uniform initial condition with the boundary values (i.e., $X(r,0)=Y(r,0)=0$ in the present case) is adopted.  (a) $D_Y=5\times10^{-6}$.  (b) $D_Y=6.4\times10^{-5}$.  (c) $D_Y=6.8\times10^{-5}$.  (d) $D_Y=1.0\times10^{-4}$.  (a) $D_Y=5\times10^{-4}$.  (a) $D_Y=9.0\times10^{-4}$. 
}
 \end{figure}

\begin{figure}
\includegraphics{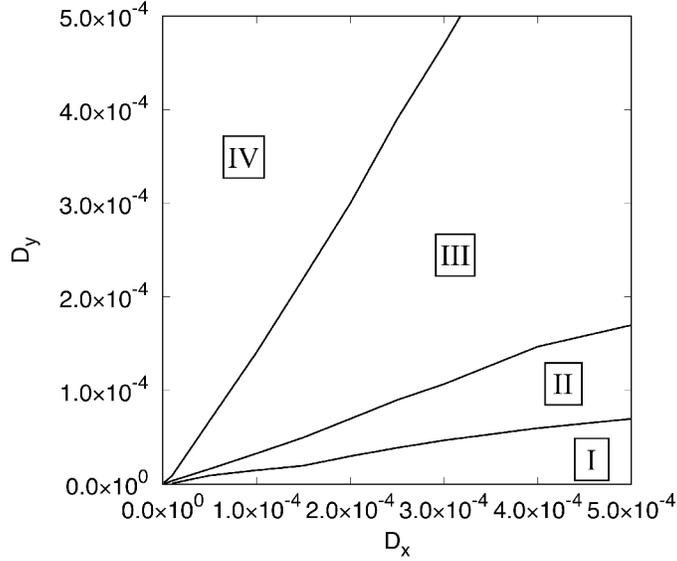}
\caption{Phase diagram of pattern formation dynamics in model \textbf{A} under a fixed boundary condition, $X_0=Y_0=0$. The abscissa is the diffusion constant of $X$, and the ordinate is that of $Y$. (Phase I): Spatially periodic and temporally fixed pattern. (Phase II): uniform oscillation except for the vicinity of fixed boundary.  (Phase III): Travelling wave. (Phase IV): Spatiotemporal aperiodic dynamics.
}
 \end{figure}
 
\begin{figure}
\includegraphics{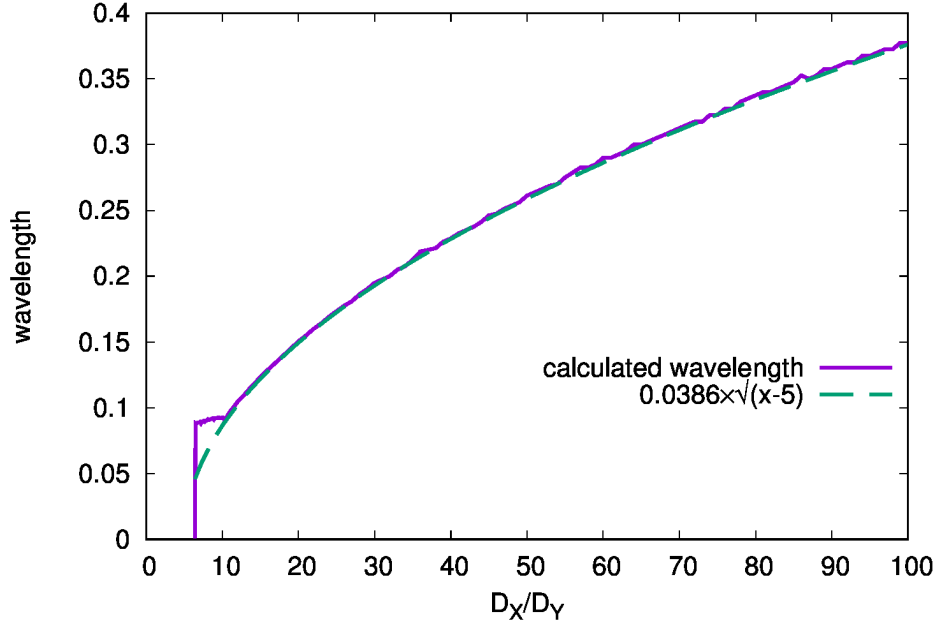}
\caption{Wavelength dependence on $D_X/D_Y$. The wavelength of fixed spatially periodic pattern is plotted against $D_X$. $D_Y$ is fixed to $1.0\times10^{-4}$. For $D_X/D_Y \lesssim 6.6$$(\sim 1/0.16)$ the uniform oscillation remains, so that a periodic pattern does not exist (See also Fig.2.). For $D_X/D_Y \gtrsim 10.5$, one stripe is formed from one period so that a trivial scaling $\lambda \propto \sqrt{D_X}$ holds. (The doted line is $\lambda \sim 0.0386 \sqrt{D_X/D_Y-5}$ for reference). For $6.6 \lesssim D_X/D_Y\lesssim10.5$, 1-to-1 locking between uniform oscillation and spatial periodic pattern collapses, and thus the wavelength is prolonged from the trivial scaling, as $n:m$ lockings ($n > 1$) appear. 
}
\end{figure}
 
\begin{figure}
\includegraphics{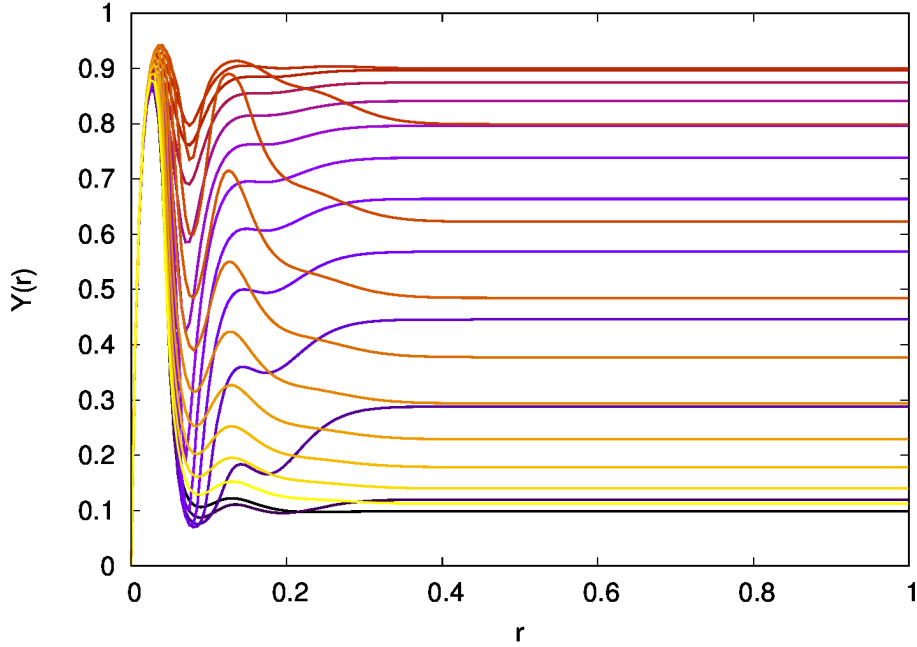}
\caption{\label{}Approach to uniform oscillatory state. $Y(r)$ at every 200 time steps from $t$ = 8800 to 10800 are superimposed with different colors. For $r \gtrsim 0.5$, spatially uniform state is reached. $D_X=5.0\times10^{-4}$ and $D_Y=1.0\times10^{-4}$, $X_0=Y_0=0$.
}
 \end{figure}
 
\begin{figure}
\includegraphics{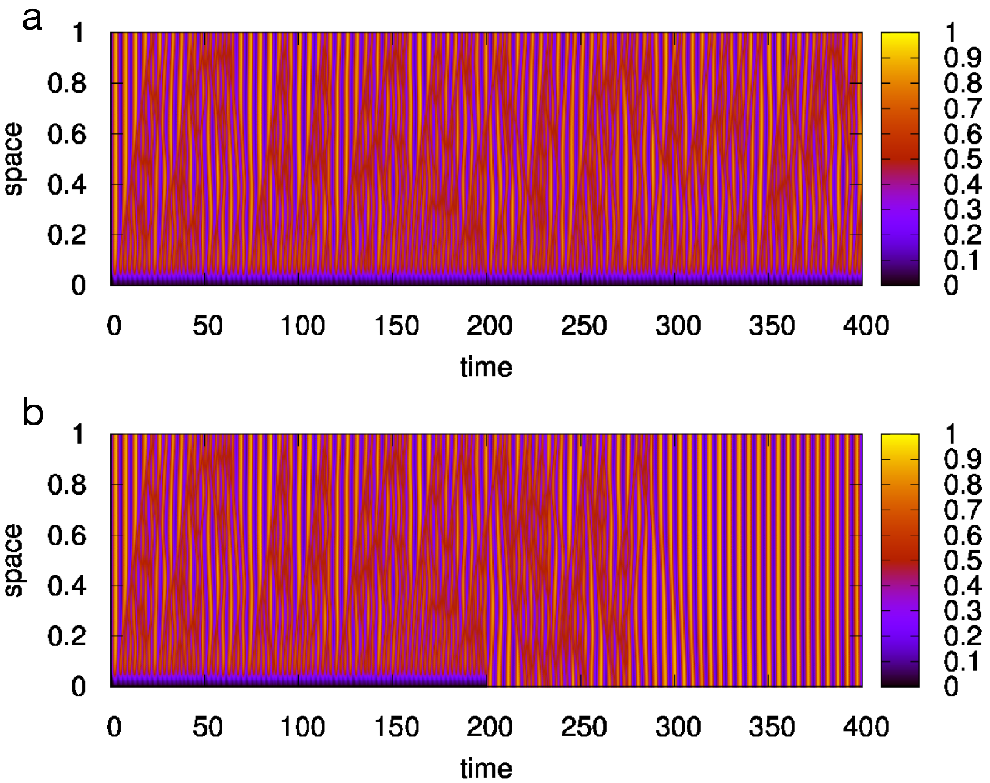}
\caption{Time development of the spatiotemporal dynamics (the dynamics in phase IV). $X(r,t)$ is displayed with a color scale given by the side bar. The abscissa is space $r$, and the ordinate is time $t$. $X_0=Y_0=0$. (a)$D_X=5.0\times10^{-4}$ and $D_Y=2.0\times10^{-3}$. (b)The boundary condition at $r=0$ is changed from $X_0=Y_0=0$(Dilichlet) to $\dfrac{\partial X}{\partial r}=\dfrac{\partial Y}{\partial r}=0$(Neumann) at $t$ = 200. After transient aperiodic spatiotemporal dynamics, uniform oscillation is attracted.
}
 \end{figure}

 Now we describe the dynamics in each phase in detail.
 \begin{enumerate}
	\item 
	When $D_Y/D_X <0.15$, i.e., the diffusion of inhibitor is much larger than that of the activator, spatially-uniform limit cycle is replaced by a spatially periodic pattern.  When $D_X$ is sufficiently larger than $D_Y$, one stripe is shaped from one periodic cycle (see Fig.1a).  Indeed this transformation from temporal oscillation to spatial pattern is recently discussed for the extreme case with $D_Y=0$, i.e., the case in which only the inhibitor diffuses \cite{kohsokabe2017boundary}.  Here, even if $D_Y$ is not zero but very small, this transformation from temporal oscillation to spatial pattern works (see the next section for the mechanism of this transformation). As $D_Y$ is increased, the formation of stripe is slowed down. One stripe is formed per two cycles (Fig.1b) and one stripe per three cycles (Fig.1c), four cycle,.., as $D_Y$ is increased. Between these simple  1:n ($n>1$) lockings complicated lockings such as 2:5 are observed as in the frequency locking within the quasiperiodic oscillation\cite{kaneko1983similarity, kaneko1986collapse, ott2002chaos}. 
	
When one stripe is formed by one periodic cycle, the proportionality between the period of the limit cycle $T$ and the wavelength of fixed pattern $\lambda$ is expected against the change in the diffusion constants (at least approximately). In this case, by using the dimension analysis\cite{kohsokabe2017boundary}, the relationship between the wavelength $\lambda$ and the oscillation period $T$ is  $\lambda \propto \sqrt{D_X} T$ when $D_Y$ is fixed at a small value.
This relationship is shown in Fig.3, against the change in diffusion constants $D_X$. Note that when $D_X/D_Y$ is decreased and 1-to-1 time-space transformation is replaced by n-to-m locking ($n>1,m\ge 1$) between cycle period and wavelength, the above proportionality is deviated to prolong the wavelength.

\item As $D_Y$ is further increased, then the diffusion of activator is sufficiently fast, so that the fixation to a spatial pattern no longer works, as the attraction to a homogeneous state is fast (Fig1d).  Near the boundary the oscillation is suppressed, but as the space goes downflow, the $(X,Y)$ values are shifted and start to oscillate. After such ``spatial transient'' the state reaches the original limit cycle with a spatially uniform state (see Fig.4). Note that in the spatial transient, slight modulation from the homogenous state remains. 

\item With the further increase in $D_Y$, the spatial transient in the upper flow is replaced by the travelling wave (see Fig.1e). The period in the wave is shorter than that of the limit cycle. Near the fixed boundary, the oscillation amplitude is small, and in the present model the oscillation period, then, is shorter than the original limit cycle.  With the diffusion of activator, this fast oscillation propagates to downflow, and hence the travelling wave is generated.

\item  With the further increase in $D_Y$, the global travelling wave turns to be unstable, and aperiodic spatiotemporal pattern replaces the uniform periodic pattern (Fig.1f).  The resultant pattern consists of combination of local travelling-wave and locally uniform periodic pattern (For the spatiotemporal pattern over much longer time scale see Fig.5a). Note the time period in the travelling wave and that of limit cycle are different and with faster diffusion of activator, aperiodic dynamics are generated accordingly.    By fixing the boundary, the present aperiodic pattern in space-time is generated, reminiscent of spatiotemporal intermittency\cite{kaneko1985spatiotemporal}.

Note that the original uniform limit cycle state remains as the attractor under a periodic or Neumann boundary condition. To examine if that is the only attractor in these boundary conditions, we change the boundary condition from fixed to Neumann condition, after the above aperiodic state in the fixed boundary is reached. Then spatiotemporal homogeneous limit cycle is reached after long transient (Fig.5b).  This transient increases rapidly with the system size, suggesting the super-transient behavior\cite{kaneko1990supertransients, crutchfield1988attractors}.

\end{enumerate}

\begin{figure}
\includegraphics{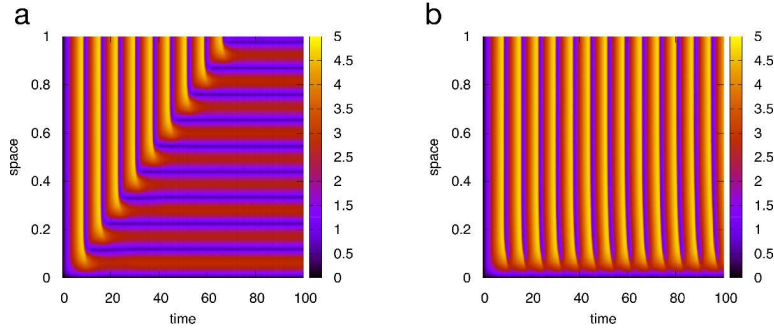}
\caption{Temporal evolution of the model \textbf{B} under a fixed boundary condition $X_0=Y_0=0$. Initial condition is an uniform state with the fixed boundary, i.e., $X(r, 0)=Y(r, 0)=0$. $X(r,t)$ is displayed with a color scale given by the side bar. The abscissa and the ordinate are time and space respectively, as in Fig.1 and Fig.5. (a) Spatially periodic and temporally stable pattern. $D_X=5.0\times10^{-4}$ and $D_Y=1.0\times10^{-5}$. (b) Uniform oscillation except for the vicinity of the fixed boundary.  $D_X=5.0\times10^{-4}$ and $D_Y=5.0\times10^{-4}$. 
}
 \end{figure}
 
 The transition between homogeneous limit cycle state and the fixed pattern is generally observed in reaction-diffusion systems. In the model A, the uniform periodic attractor remains to exist near the Hopf bifurcation point $\beta = 8$, but for $\beta \gtrsim 12$, the fixation to spatially periodic state is observed. In Fig.6, we have plotted the case with model B. Again one stripe is generated from one cycle when $D_X\gg D_Y$. In this case, however, we have observed only the phases I and II.

\section{Transformation from temporal oscillation to spatial pattern: explanation by nullcline shift}

Now we analyze how oscillation is fixed into a periodic spatial pattern.
Note that a uniform oscillatory state exists for $r>0$ initially.
Thus the pattern formation progresses far from a fixed point state. Accordingly, the standard analysis for Turing instability cannot be applied, since it treats the instability of a spatially-uniform and temporally fixed point state against perturbation.
 
\begin{figure}
\includegraphics{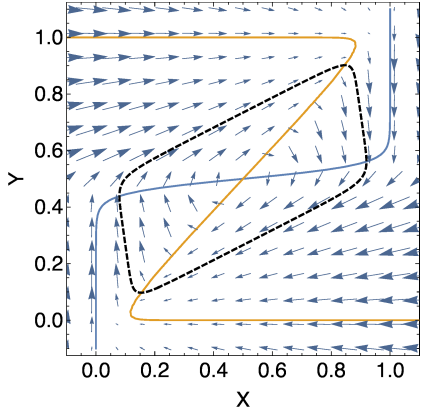}
\caption{The nullclines $f(X,Y)=0$ (blue) and $g(X,Y)=0$ (orange) are plotted in the state space for the model (\textbf{A}). The abscissa is $X$, and the ordinate is $Y$. Flow in the state space is also illustrated by arrows.
Limit cycle attractor is also plotted by the broken curve.
}
 \end{figure}
  
Intuitively, the fixation mechanism we found is understood as follows. In a homogeneous state, the temporal evolution is given by the local dynamics $dX/dt=f(X,Y),dY/dt=g(X,Y)$. There are no stable fixed points given by the crossing point of $X-$ and $Y-$ nullclines (see Fig.7).  Due to the fixed boundary, however, oscillation stops at $r\sim 0$, and there, the values $X,Y$ cannot remain identical with those at the limit cycle attractor.  Then the diffusion gives an additional term to each of the local dynamics for a downflow point $r\gtrsim0$. Accordingly, the nullclines for the local dynamics including such diffusion terms could be shifted from the original, so that they can cross to generate a novel stable fixed point. Thus a fixed point replaces the oscillatory state at $ r \sim 0$. This gives a spatial gradient again to further downflow, so that diffusion term gives shift of nullclines for the local dynamics at the downflow.  If this shift is sufficient to make the nullclines cross with each other, a fixed point is again generated at the downflow. With the repetition of this process, the oscillatory state is replaced by a fixed spatial pattern.

\begin{figure}
\includegraphics{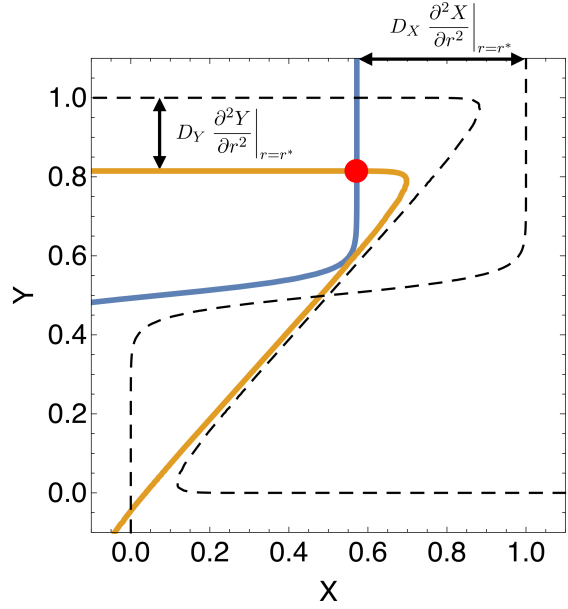}
\caption{State-space plot of the local nullcline of $X$ (blue solid) and the nullcline of $Y$ (orange solid) of the stationary state resulting from the dynamics in the phase (I), at some location $r^*$. The abscissa is $X$, and the ordinate is $Y$. The two dashed lines are the original nullclines, i.e., $f(X,Y)= 0$ and $g(X,Y) = 0$ (see also Fig.7). Distance between each local nullcline and the original nullcline is provided by the local diffusion terms, i.e., $D_{X}\left.\dfrac{\partial^2 X}{\partial r^2}\right|_{r=r^*}$ and   $D_{Y}\left.\dfrac{\partial^2 Y}{\partial r^2}\right|_{r=r^*}$. The nullcline of $X$ is shifted horizontally and the nullcline of $Y$ is shifted diagonally. The stationary state from the initial condition $X(r,0)=Y(r,0)=0$ is also depicted as the red circle, which lies at the cross point of two nullclines.
}
 \end{figure}
 
Now we discuss if this scenario really works in the reaction-diffusion systems we studied here. The fixed point pattern should satisfy 
\begin{equation}
\left.
\begin{array}{l}
f(X(r),Y(r))+D_X\dfrac{\partial^2 X}{\partial r^2}=0\\[12pt]
g(X(r),Y(r)) +D_Y\dfrac{\partial^2 Y}{\partial r^2}=0\\[10pt]
\end{array}
\right.
\label{eq:stationaryequation}
\end{equation}
Eq.(\ref{eq:stationaryequation}) describes local nullclines of $X$ and $Y$ for the fixed pattern, at each location $r$. Now, consider the fixed boundary at $r=0$. In the model (A), the term $ D_X\dfrac{\partial^2 X}{\partial r^2}$ at $r>0$ gives a horizontal shift in X-nullcline, and  the term $ D_Y\dfrac{\partial^2 Y}{\partial r^2}$ gives a diagonal shift in Y-nullcline. In Fig.8, we plotted how these nullclines are shifted. With this shift, a pair of cross-points is generated, one of which give a stable fixed point. The diffusion term successively allows for the emergence of a stable fixed point that varies with $r$.

	To sum up, a stable fixed point is generated by the shift of nullclines, while the chain of such novel fixed points along space generate the diffusion term which self-consistently generates the shift.

\begin{figure}
\includegraphics{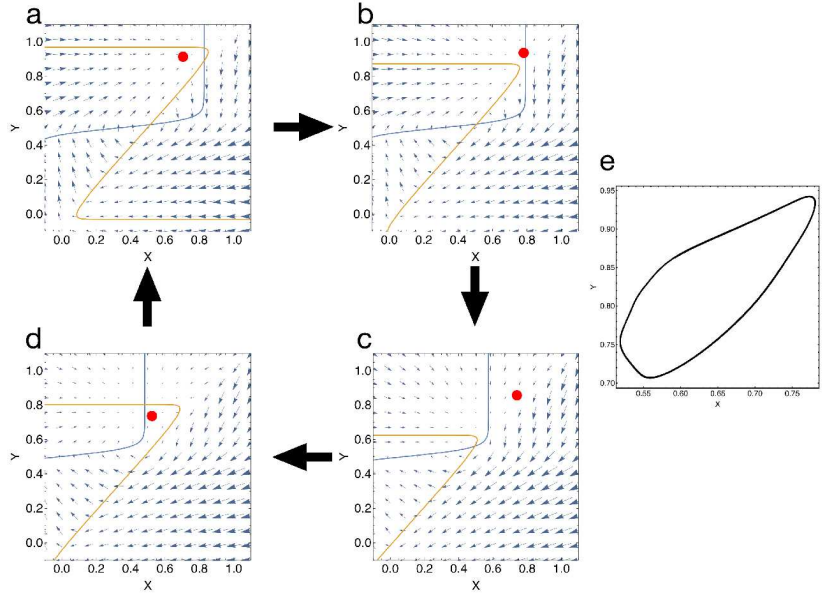}
\caption{(a-d) Temporal change of the local nullclines shifted by the diffusion term and the state values, in the phase (II). For different time points within one period, the local nullclines of $X$ (blue) and $Y$ (orange) and the state value ($X,Y)$ at the moment (red circle) are plotted successively at (a)-(d), for a position in the vicinity of the fixed boundary, where the uniform oscillation is not yet reached (see Fig.5). The abscissa and the ordinate are $X$ and $Y$  as in Fig.7 and Fig.8. (a) At a certain time, a pair of cross points of nullclines emerges, when the state value is directed to one of them, but has not yet reached the stable fixed point. (b)As the transient state approaches the stable fixed point, local net diffusion changes accordingly, and then the pair of cross-points of nullclines, and consequently the stable fixed point, disappear. (c, d)Cross-points are not generated except the original unstable fixed point, the source of original limit cycle attractor. Hence the state rotates around the unstable fixed point, until it returns to (a). Overall, the orbit of the transient dynamics forms limit cycle attractor, as is shown in (e).
}
 \end{figure}

Now, we study how the transition to the homogeneous limit cycle state occurs in terms of nullcline shift. In the above case, as $D_X\gg D_Y$, the horizontal shift of X-nullcline is dominant, which keeps on generating a pair of novel crossing points of the nullclines, one of which corresponds to a stable fixed point. As the diffusion of activator is increased, however, such cross-point is not sustained with time, as shown in Fig.9. Hence the temporal oscillation remains even at upper flow.  Further downflow, the spatial gradient in $X(r)$ or $Y(r)$ decreases, and such small shift can no longer generate a crossing point. At further down flow, the spatial uniformity is reached, and no more shifts of nullclines exist, so that the original flow in Fig.7 is recovered, giving rise to the original uniform limit cycle attractor (see also Fig.5).

	With the further increase in $D_Y$, the diagonal shift of nullclines is larger. Again, crossing-points of nullclines appear at a certain moment at the upper flow.  The actual value of $(X(r,t),Y(r,t))$, however, is deviated from the crossing point, so that the self-consistent condition for the crossing point and the actual diffusion term to give the shift is not satisfied.  In the travelling wave phase (III), however, even though $(X(r,t),Y(r,t))$ is no longer fixed in time for given $r$, the solution written as $(X(R),Y(R))$ with $R=r-vt$ and $v$ as a speed of travelling wave. Then, instead of the condition $\partial X/\partial t=f+D_X\partial^2X/\partial r^2=0$ to be satisfied for the fixed pattern, $-vdX/dR=f+D_X\partial^2X/dR^2$ has to be satisfied ($Y$ has to satisfy the corresponding equation). In other words, even though the shifted nullclines do not keep on generating a requested crossing point, instead, the fixed point condition for nullcline crossing is expected to be satisfied in an appropriate inertial frame that moves with the speed of the traveling wave.

	As $D_Y$ is further increased, $X$- and $Y$-nullclines can cross at some spatial location but do not remain crossing globally over the space. No moving frame can keep the crossing points of nullclines, either.  This leads to the aperiodic spatiotemporal dynamics.

For the model (B) again, the shifts of $X$- and $Y$- nullclines by the diffusion term generate a crossing point and accordingly a fixed pattern is formed.

\section{Spatial Map}

When the fixed pattern exists, $\partial X/\partial t=0$ and $\partial Y/\partial t=0$ are satisfied. 
This condition corresponds to the self-consistent shift of nullclines, and with this condition, the pattern is successively determined from the boundary point.
To formulate this fixed-pattern condition, we adopt spatial discretization of the reaction-diffusion equation. Then the condition for a fixed spatial pattern is given by 
\begin{equation}
\left.
\begin{array}{l}
f(X_{\ell},Y_{\ell})+D_X^{'}(X_{\ell+1}-2 X_{\ell}+X_{\ell-1})=0\\
g(X_{\ell},Y_{\ell})+D_Y^{'}(Y_{\ell+1}-2 Y_{\ell}+Y_{\ell-1})=0\\[10pt]
\end{array}
\right.
\label{eq:discretized_equation}
\end{equation}
where $\ell$ is a discretized space index with $a$ as the discretized unit length, i.e., $r=a\ell$ and $X_{\ell}=X(r/a)$, $Y_{\ell}=Y(r/a)$. Here, $D_{X,Y}^{'}$ is the rescaled diffusion constant which satisfies $D_{X,Y}{'}={D_{X,Y}}/{a^2}$. From these equations, we can derive the four-dimensional map as follows:
\begin{equation}
\left.
\begin{array}{l}
X_{\ell+1}=2X_{\ell}-X_{\ell-1}-\dfrac{f(X_{\ell}, Y_{\ell})}{D_X^{'}}\\[12pt] 
Y_{\ell+1}=2Y_{\ell}-Y_{\ell-1}-\dfrac{g(X_{\ell}, Y_{\ell})}{D_Y^{'}}\\
\end{array}
\right.
\label{eq:spatialmap}
\end{equation}

Previously we considered the system with $D_Y=0$, in which case the mapping is reduced to a two-dimensional map $(X_{\ell-1},X_{\ell})\rightarrow X_{\ell+1}$, as $Y_{\ell}$ is determined from $g(X_{\ell}.Y_{\ell})=0$\cite{kohsokabe2017boundary}.  This two-dimensional map is more tractable than the above four-dimensional mapping, but instead, additional selection criterion was needed to choose an appropriate branch in the mapping in obtaining $Y_{\ell}$ from $g(X_{\ell}.Y_{\ell})=0$ (The use of a ``spatial map'' was introduced earlier in the analysis of a spatial pattern, see also \cite{aubry1978new, willeboordse1995pattern}). 
	
From this four-dimensional mapping, the spatial pattern is provided as a time-series generated by the mapping (\ref{eq:spatialmap}), and then by replacing $\ell$ to spatial sequence $r=a\ell$, the spatial pattern is obtained. 

\begin{figure}
\includegraphics{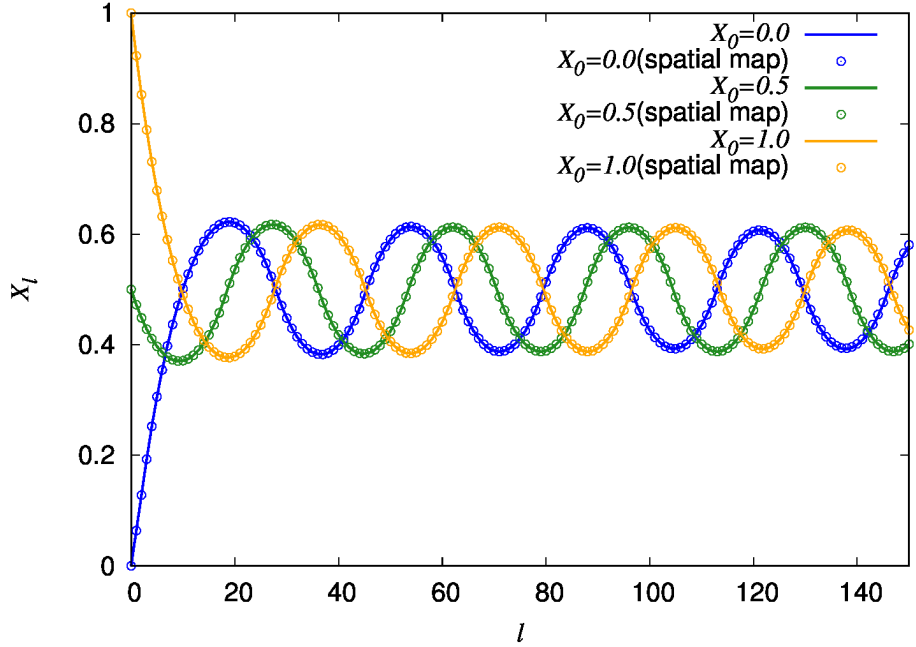}
\caption{Stationary pattern $(X(r), Y(r))$ of the model (\textbf{A}) and the predicted pattern of $(X_l, Y_l)$ obtained from the spatial map (\ref{eq:spatialmap}). The predicted pattern (given by $\circ$) agrees well with the stationary pattern (lines) obtained from the direct simulation of the model (\textbf{A}). Three patterns with boundary values of $X_0 =Y_0=0.0$ (blue), $X_0 = 0.5$ and $Y_0=0.0$ (green), and $X_0=1.0$ and $Y_0=0.0$ (orange) are plotted, while for direct simulations, the uniform initial condition with the corresponding boundary values is taken.  $D_X=5.0\times10^{-4}$ and $D_Y = 5.0\times10^{-5}$.
}
\end{figure}

Examples of spatial sequences derived from Eq.7, for given three different initial conditions are plotted in Fig.10, which reproduces the fixed patterns obtained from direct numerical simulation of the reaction diffusion equation. Now note that the spatially periodic pattern is obtained as an attractor of the spatial map (\ref{eq:spatialmap}), 
hence, independent of the value of the fixed boundary, the periodic pattern of the same wavelength (only with the phase difference) is obtained, as is observed in the direct simulation.
 (see Fig. 10).

As $D_Y$ is increased, the spatial period increases from 1 to 2, and 3 (see Fig.1c,d). With further increase in $D_Y$, the periodic attractor in the spatial map disappears, and the fixed pattern is no longer generated in the original equation (1), either. 

Also, when $\beta$ in the model (A) is decreased, the pattern predicted by the spatial map can be quasiperiodic or chaotic, and again, a fixed spatial pattern is not generated, and is replaced by aperiodic spatiotemporal dynamics.

Finally, the spatial map works well in the model (B) also.

\section{Discussion}

	In the present paper, we have studied the boundary-condition-induced pattern dynamics from a uniform oscillatory state in a reaction-diffusion system. By imposing a fixed boundary condition, the spatially-uniform limit cycle state is  replaced by the following three phases, depending on the diffusion constants, : (I) fixed stripe pattern (III) travelling wave from the boundary (IV) aperiodic spatiotemporal pattern with the mixture of local travelling-wave and locally uniform oscillatory states, instead of the (II) original spatially-uniform limit cycle state. Note that in all cases, uniform oscillatory states are stable as long as the adopted boundary condition allows for such state globally, as given by, say, a periodic or Neumann boundary condition. In contrast, just by fixing a value at the boundary, the uniform state is replaced by the spatiotemporal patterns (I), (III) and (IV).

     To analyze how the limit cycle is transformed to spatial pattern in the case (I), we introduced a diffusion term into the (local) nullcline analysis at each spatial point. Each diffusion term gives a shift for each nullcline, by which a novel crossing-point of nullclines is generated, giving rise to a stable fixed point. If the chain of values of the generated stable fixeds point is consistent with the diffusion term needed for the each corresponding nullcline shift, then the fixed spatial pattern is stable and replaces the homogeneous state.  The transitions among the above phases (I)-(IV) are also analyzed with this analysis of nullcline shift. Further, the fixed spatial pattern implies a sequence of conditions to be satisfied from the upper fixed boundary. These conditions are represented by a four-dimensional mapping from upper to down flow in space. The attractor of the mapping, by re-interpreting the time as space, determines the spatial pattern.

The present mechanism of boundary-induced pattern formation is expected to work in a variety of systems that have a spatially uniform limit cycle attractor when the values of diffusion constants are appropriate. As such limit cycle attractor is common in a two-variable system with a negative-feedback interaction, say with an activator and an inhibitor, there will be a variety of systems that suit the present mechanism. Also, extension to a two or three-dimensional space is straightforward, where patterns are formed depending on the conditions on boundaries. 

Further, the boundary-condition dependence, which is not fully explored as compared with the initial-condition dependence, will be important in a broad area of nonlinear systems. By using the analysis by spatial map, the spatial pattern is obtained by the evolution of the mapping, and the initial condition in the mapping corresponds to the boundary condition. Hence the standard analysis of dynamical systems is applicable. The spatial map will provide a useful tool to study the boundary-induced formation of fixed patterns.

Considering that the real system in nature is always finite in space, the boundary-condition can influence the system's behavior. The present result implies that the boundary can influence not only on its vicinity but globally throughout the space. Besides the stripe formation, spatiotemporal aperiodic behavior (chaos) is generated from a stable period oscillation just by fixing the boundary. It will be interesting to examine experimentally the observed behaviors in reaction-diffusion and other spatially extended systems.

   Note that we first found the present fixation of oscillation in the evolution simulation to fit the protein expression pattern generated by gene-regulation-network with a requested spatial pattern\cite{kohsokabe2016evolution}. Instead of the standard Turing pattern, we observed the present oscillation-fixation mechanism frequently in these evolution simulations. Furthermore, in biological morphogenesis, oscillatory gene expressions are often observed, while the boundary condition is essential to morphogenesis. 

In this context, it is interesting to recall that oscillation in protein expression is observed in somitogenesis \cite{palmeirim1998uncoupling,pourquie2001vertebrate} and in neural and other stem cells\cite{shimojo2008oscillations, kobayashi2009cyclic}. In the somitogenesis, such oscillation is transformed into spatial pattern, where relevance of cell-cell interaction has recently been discussed\cite{cotterell2015local}. Also under {\sl in vitro} condition, the differentiation from such oscillation is recently observed experimentally, where both the gradient due to cellular heterogeneity and cell-cell interaction seem to be important to induce the differentiation from the oscillatory dynamics \cite{aulehla2004segmentation, tsiairis2016self}. In the present study, the fixed boundary generates a gradient, which causes the fixed pattern with differentiation of protein expressions $X$ and $Y$ through the diffusive interaction, as is consistent with the experiment. Further comparison with biological experiments on morophogenesis will be important both in biology and in nonlinear dynamics.

The present manuscript is dedicated to the memory of John Hudson. I (K.K.) first read his paper on chaos in BZ reaction\cite{hudson1979experimental}, about 36 years ago, and since then I have always been impressed by his marvelous studies  unveiling beautiful structure from chemical-reaction experiments that could be potentially complicated. It was my fortune that our studies come closer, through his beautiful experiment on clustering\cite{wang2000experiments} in electrochemical reaction which exhibits amazing correspondence with the behaviors in globally coupled maps\cite{kaneko1990clustering}. I regret that I could no longer hear his invaluable opinion on the present study, based on his deep insight as experimentalist.

\begin{acknowledgments}
The authors would like to thank Tetsuhiro Hatakeyama for stimulating discussions.
This research was partially supported by a Grant-in-Aid for Scientific Research (S) (15H05746) from the Japanese Society for the Promotion of Science (JSPS) and Grant-in-Aid for Scientific Research on Innovative Areas (17H06386) from  the Ministry of Education, Culture, Sports, Science and Technology (MEXT) of Japan.
\end{acknowledgments}

%

\end{document}